\documentclass {JHEP3}

\def\IR{\relax{\rm I\kern-.18em R}}

\def\CF{{\cal F}}
\def\CA{{\cal A}}
\def\CL{{\cal L}}
\def\CM{{\cal M}}
\def\CK{{\cal K}}

\def\IR{\relax{\rm I\kern-.18em R}}

\def\CF{{\cal F}}
\def\CA{{\cal A}}
\def\CL{{\cal L}}
\def\CM{{\cal M}}
\def\CK{{\cal K}}

\author{Aybike \c{C}atal-\"{O}zer \\
{\it School of Mathematics, Trinity College Dublin\\
Dublin 2 IRELAND \\ \underline{e-mail}: aybike@maths.tcd.ie} }

\abstract{We study duality-twisted dimensional reductions on a
group manifold G, where the twist is in a group $\tilde{G}$ and
examine the conditions for consistency.  We find that if the
duality twist is introduced through a group element $\tilde{g}$ in
$ \tilde{G}$, then the flat $\tilde{G}$-connection $A =
\tilde{g}^{-1} d\tilde{g}$  must have constant components $M_n$
with respect to the basis 1-forms on $G$, so that the dependence
on the internal coordinates cancels out in the lower dimensional
theory. This condition can be satisfied if and only if $M_n$ forms
a representation of the Lie algebra of $G$, which then ensures
that the lower dimensional gauge algebra closes. We find the form
of this gauge algebra and compare it to that arising from flux
compactifications on twisted tori. As an example of our
construction, we find a new five dimensional gauged, massive
supergravity theory by dimensionally reducing the eight
dimensional Type II supergravity on a three dimensional
unimodular, non-semi-simple, non-abelian group manifold with an
$SL(3,\IR)$ twist. }

\title{DUALITY TWISTS ON A GROUP MANIFOLD}

\begin{document}


\section{Introduction}

In the toroidal dimensional reduction of a theory invariant under
a global symmetry group G, it is possible to introduce a
generalized ansatz for the reduction of the fields transforming in
a non-trivial representation of G. The ansatz which was fist
introduced by Scherk and Schwarz in \cite{SS1} is
\begin{equation}\label{SSansatz}
\hat{\phi}(x^{\mu},y^m) = g(y^m) \phi(x^{\mu})
\end{equation}
where $\hat{\phi}$ is a generic field transforming under G as
$\hat{\phi} \rightarrow g \hat{\phi}$ and $y^m, m=1,\cdots,d$ are
coordinates on $T^d$ so that $g : T^d \rightarrow$ G. The ansatz
(\ref{SSansatz}) is equivalent to an expansion of the fields in
terms of the harmonics of $T^d$ followed by a consistent
truncation to the zero modes with a twisted boundary condition for
$\hat{\phi}$ (as opposed to the periodic boundary conditions
imposed by the standard Kaluza-Klein ansatz). As $\hat{\phi}$
traverses a cycle of $T^d$ parameterized by $0 \leq \tau_i \leq 2
\pi R_i, i=1, \cdots,d$ it picks up a monodromy $\Omega_i(g)$ so
that the twisted boundary condition is
\begin{equation}\label{bc}
\hat{\phi}(x^{\mu}, \tau_i = 2 \pi R_i) = \Omega_i \
\hat{\phi}(x^{\mu}, \tau_i = 0).
\end{equation}
The monodromies introduce in lower dimensions a non-abelian gauge
algebra, mass parameters and a scalar potential. The G-invariance
of the higher dimensional theory ensures that the $y$-dependence
cancels out in the lower dimensional action and the reduction is
consistent in the sense that the solutions to the lower
dimensional field equations can be uplifted to become solutions of
the higher dimensional ones. In the recent literature, such
dimensional reductions are usually called reductions with duality
twists \cite{Dabholkar:2002sy}.

Later in \cite{SS2}, Scherk and Schwarz introduced a related
scheme of generalized dimensional reduction, where now the global
symmetry exploited is `internal' as opposed to the `external'
symmetries of \cite{SS1}. In the terminology of \cite{SS1} and
\cite{SS2}, the internal symmetries are the geometric symmetries
associated with the internal manifold, while the external
symmetries act on the spinor and p-form fields. The scheme
introduced in \cite{SS2} can be described as the dimensional
reduction on a d-dimensional parallelizable manifold X with
well-defined nowhere vanishing basis one-forms
\begin{equation}\label{vielbein}
\eta^m = U^m_{\ \ n}(y) dy^n, \end{equation} where $y^m$ are
coordinates on X and $U^m_{\ n}(y)$ is a matrix element of the
internal symmetry group G. The one-forms $\eta^m$ satisfy
\begin{equation}\label{structure}
d\eta^m + \frac{1}{2} C^m_{\ np} \eta^n \wedge \eta^p = 0
\end{equation}
with  coefficients
\begin{equation}\label{constant}
C^m_{\ np} = - 2(U^{-1})^r_{\ n}(U^{-1})^s_{\
p}\partial_{[r}U^m_{\ \ s]}.
\end{equation}
Consistency requires $C^m_{\ np}$ to be constant, which in turn
implies that they are  the structure constants of the Lie algebra
of G. Then locally the internal space X has the structure of the
group manifold of G. Globally X = G/$\Gamma$ where $\Gamma$ is a
discrete subgroup of G \cite{Hull:2005hk} and hence its structure
can be quite different from the group manifold. It is common in
the literature to refer to such reductions as Scherk-Schwarz
reductions and the internal space X as the twisted torus. In this
paper, we are mainly interested in the local structure, so for our
purposes here a twisted torus is a group manifold, where the group
can be and in general is non-compact. Like with the reductions
with duality twists, reductions on twisted tori too introduce in
lower dimensions a non-abelian gauge algebra, mass parameters and
a scalar potential.

As was already mentioned in \cite{SS1} and \cite{SS2}, in some
cases dimensional reduction on a twisted torus can be equivalent
to a standard Kaluza-Klein reduction followed by a dimensional
reduction with a duality twist. Indeed, after a Kaluza-Klein
reduction, the internal symmetries associated with the geometry of
the internal manifold are promoted to the external symmetries of
the lower dimensional theory, which then can be exploited in a
subsequent reduction with a duality twist. For example, consider a
theory compactified on a two torus $T^2$. The lower dimensional
theory has an $SL(2,\IR)$ symmetry as part of its global symmetry
group, as $SL(2,\IR)$ is the large diffeomorphism group of $T^2$.
In the spectrum of the theory there exists two scalar fields,
$\tau_1$ and $\tau_2$, which correspond to the moduli
parameterizing the shape of the internal $T^2$, transforming under
$SL(2,\IR)$ through fractional linear transformations. Now, in a
further compactification on a circle $S^1$ we can introduce a
duality-twisted ansatz for these fields as in (\ref{SSansatz}),
where $g(y)$ is in $SL(2,\IR)$. From the point of view of the
parent theory, this is nothing but a compactification on a three
dimensional twisted torus with the metric
\begin{eqnarray}\label{torusmetric}
 ds^2 &=& (2 \pi R)^2 dy^2 + \frac{A}{\tau_2(y)} \mid \tau(y)
 dx_1 + dx_2  \mid^2,
\end{eqnarray}
from which we can check that the basis 1-forms $\eta^m$ satisfy
(\ref{structure})\footnote{Duality-twisted reductions are
classified with respect to the conjugacy classes of the duality
group. In each conjugacy class of $SL(2,\IR)$ a representative can
be chosen such that $C^m_{\ np}$ in (\ref{structure}) are
constants. See section 4.}. Here $y$ parameterizes the $S^1$, $R$
is the radius of $S^1$, $A$ is the area of $T^2$ and $\tau =
\tau_1 + i \tau_2$. We will discuss this in more detail in section
four.

The purpose of this paper is to consider twisted dimensional
reductions on a group manifold G, where the twist is in a duality
group $\tilde{G}$ and  examine the conditions for which the ansatz
(\ref{SSansatz}) yields a consistent dimensional reduction. We
find that the Lie algebra (of $\tilde{G}$) valued  one-form $A =
\tilde{g}^{-1} d\tilde{g} = A_m(y) \eta^m$ (with $\tilde{g} \in
\tilde{G}$) should have constant components $A_n(y) = M_n$ so that
the $y$-dependence cancels out in the lower dimensional action and
the field equations.  As soon as we impose the condition that
$M_n$ must be constant elements of the Lie algebra of $\tilde{G}$,
not depending on the coordinates $y^m$ of $G$, we see that $M_n$
must also satisfy the following commutation relations
\begin{equation}\label{comm1}
[M_n, M_p] = C^{q}_{\ \ np} M_q.
\end{equation}
This follows from the fact that the 1-form $A$, being of the form
$A=\tilde{g}^{-1}d\tilde{g}$ satisfies the zero curvature
condition
\begin{equation}\label{flat}
    dA + A \wedge A = 0.
\end{equation}
We also find, the condition (\ref{comm1}) ensures the closure of
the  lower dimensional gauge algebra arising from the
$\tilde{G}$-twisted reduction on $G$. At this point, an important
question arises as to whether $A$ is pure gauge globally or only
locally. If $\tilde{g}$ is single-valued on $G$ so that $A$ is
pure gauge globally, then $A$ can be gauge transformed to a zero
connection, rendering our dimensional reduction equivalent to a
standard group manifold reduction on $G$, with no $\tilde{G}$
twist at all. However, if $G$ is not simply connected with
$\pi_1(G) \neq 0$, then one can introduce non-trivial monodromies
for the connection $A$ over the cycles of $G$, which then
introduces twisted boundary conditions for the fields charged
under the duality group $\tilde{G}$. In this case, $M_n$
introduces as usual the mass terms and the gauge parameters in the
lower dimensional theory.

The plan of the paper is as follows. In the next section we
briefly review the standard dimensional reduction on a group
manifold. In section 3 we study the twisted dimensional reduction
on a group manifold G of a particular $\tilde{G}$-invariant theory
of gravity coupled to scalars and p-form fields. We find the
consistency conditions for the cancellation of the $y$-dependence
and the closure of the lower dimensional gauge algebra. In section
4 we turn to our main interest: unimodular, non-semi-simple group
manifolds of dimension three. We review in this section that all
such manifolds are locally isomorphic to a twisted torus with the
metric (\ref{torusmetric}), where $\tau(y)$ is given by
(\ref{SSansatz}) with $g(y)$ in a certain conjugacy class of
SL(2). In section 5, we consider the low energy effective field
theory of eight dimensional type II string theory. In this
dimension, the U-duality group is SL(2) $\times$ SL(3).
Dimensionally reducing on a three dimensional unimodular,
non-semi-simple group manifold G with an SL(3) twist, we obtain in
five dimensions a new gauged supergravity with mass terms and a
scalar potential. We conclude with discussions in section 6.


\section{Dimensional Reduction on a Group Manifold}
The metric ansatz which leads to a consistent dimensional
reduction from $D+d$ to $D$ dimensions on a $d$-dimensional group
manifold G is the following
\begin{equation}\label{gmfd}
\hat{ds}^2 = e^{2\alpha \phi}ds^2 + e^{2\beta \phi}\CM_{mn}(\eta^m
+ \CA^m_{\mu}dx^{\mu})(\eta^n + \CA^n_{\mu}dx^{\mu}).
\end{equation}
Here $y^m$ are the coordinates on the group manifold, which
consists of the group elements $g(y^m) = g \in G$. That is, we
have a  group element of $G$ corresponding to each point on the
group manifold. The $\eta^m(y) = \eta^m_{\ n}(y)dy^n$ with
$\eta^m_{\ n}(y) \in G$ are the basis 1-forms on $G$, $ds^2$ is
the metric on the $D$ dimensional space-time, $\phi$ is the
dilaton and the vectors $\CA^n = \CA^n_{\mu} dx^{\mu}$ are the $d$
graviphotons. $\CM$ is a scalar matrix parameterizing the coset
$SL(d,\IR)/SO(d)$ corresponding to $d-1$ dilatons and $d(d-1)/2$
axions. It is convenient to set the values of $\alpha$ and $\beta$
to
\begin{equation}\label{alfa}
\alpha^2 = \frac{d}{2(D+d-2)(D-2)}, \ \ \ \ \beta =
-\frac{(D-2)\alpha}{d}
\end{equation}
so that the lower dimensional Einstein-Hilbert action has the
conventional form. The internal part of (\ref{gmfd}) corresponds
to the left invariant metric of a group manifold
\begin{equation}
ds^2_G = e^{2\beta\phi} \CM_{mn} \eta^m \eta^n.
\end{equation}
In order for the lower dimensional theory to be independent of the
internal coordinates, the internal dependence of $\eta^m(y) =
\eta^m_{\ n}(y)dy^n$ should be chosen such that
\begin{equation}
t_m \eta^m_{\ n} dy^n = g^{-1} dg
\end{equation}
for group elements $g = g(y^m)$, where $t_m$ forms a basis for the
Lie algebra of $G$. As a result, $\eta^m$ are the left invariant
Maurer-Cartan forms of a group manifold G, satisfying the
condition (\ref{structure}) with constant structure constants
$C^m_{\ np}$. The metric ansatz (\ref{gmfd}) yields a consistent
reduction  of the $D+d$ dimensional Einstein-Hilbert action to $D$
dimensions, provided that the group $G$ is
unimodular\footnote{This means that the adjoint representation of
the group has unit determinant. At the level of the Lie algebra,
this implies that the structure constants are traceless, i.e,
$C^m_{\ mn} = 0$. Equivalently, any left-invariant measure on G is
also right-invariant, so all top-dimensional forms on $G$ are
proportional up to a factor which does not depend on the
coordinates on $G$. All compact groups and semi-simple groups
(compact or not) are necessarily unimodular \cite{miriam}.}. If
$G$ is not unimodular, then there is a consistent reduction at
level of field equations only \cite{miriam,Roest:2004pk}.

From the dimensional reduction of the $D+d$ dimensional
Einstein-Hilbert action on a $d$ dimensional unimodular group $G$,
one obtains the following Lagrangian in $D$ dimensions
\begin{equation}\label{EHgmfd}
\CL_{EH} = R * 1 + \frac{1}{4}{\rm Tr}(D\CM \wedge * D\CM^{-1}) -
\frac{1}{2}d\phi \wedge * d\phi - \frac{1}{4}e^{2(\alpha
-\beta)\phi}\CF^m\CM_{mn} \wedge * \CF^n - V.
\end{equation}
Here
\begin{equation}\label{F}
\CF^m = d\CA^m - \frac{1}{2}C^m_{\ np}\CA^n \wedge \CA^p,
\end{equation}
\begin{equation}\label{M}
D\CM_{mn} = d\CM_{mn} + 2C^p_{\ q(m}\CA^q\CM_{n)p}
\end{equation}
and the scalar potential $V$ is
\begin{equation}\label{potential}
V = \frac{1}{4}e^{-2(\alpha -\beta)\phi}[2\CM^{nq}C^p_{\ mn}C^m_{\
pq}+\CM^{mq}\CM^{nr}\CM_{ps}C^p_{\ mn}C^s_{\ qr}].
\end{equation}

If the $D+d$ dimensional theory also includes p-form fields
$B_{(p)}$, the corresponding ansatz for their reduction is
\begin{eqnarray}\label{pform}
\hat{B}_{(p)} &=& B_{(p)}+ B_{(p-1)m} \wedge h^m + \cdots +
\frac{1}{(p-k)!} B_{(k)m_1\cdots m_{p-k}} \wedge h^{m_1} \wedge
\cdots \wedge h^{m_{p-k}}.
\end{eqnarray}
Here $h^m = \eta^m + \CA^m$ and $k$ is the larger of zero and
$p-d$. The hatted fields on the left hand side are $D+d$
dimensional, whereas the unhatted fields $B_{(q)}$ on the right
hand side are $D$ dimensional and do not depend on the $d$
internal coordinates for the consistency of the ansatz. For more
details on group manifold reductions, see for example
\cite{miriam,Roest:2004pk}.


\section{Twisted Reductions on a Group Manifold}
In the toroidal reduction of a $\tilde{G}$-invariant theory with a
duality twist with an ansatz of the form (\ref{SSansatz}), an
important consistency criterium is that
$\tilde{g}(y)^{-1}d\tilde{g}(y)$ should be a constant 1-form. This
condition ensures that the dependence on the internal coordinates
cancels out in the lower dimensional action and field equations.
When the internal space is a circle an obvious consistent choice
is $\tilde{g}(y) = e^{My}$, with $M$ in the Lie algebra of
$\tilde{G}$. For the generalization to a $d$ dimensional torus
$T^d$ the appropriate choice is
\begin{equation}\label{g2}
\tilde{g}(y^1, \cdots, y^n) = e^{M_1 y^1 + \cdots + M_d y^d},
\end{equation}
where $y^m$ are coordinates on $T^d$. The matrices $M_n$ in
(\ref{g2}) are required to commute, so that the $y$ dependence
cancels out and the gauge algebra closes in the lower dimensional
theory. In this section we examine the case in which the internal
space is a group manifold $G$. For this purpose, we study a
particular type of Lagrangian for simplicity. Namely, we consider
the group manifold reduction of a theory of gravity coupled to
scalars in the coset $\tilde{G}/H$ (where $H$ is the maximally
compact subgroup of $\tilde{G}$) and a set of $r$ {} $n-1$ form
gauge potentials $\hat{B}_{(p)}^a$ with $n$-form field strengths
$\hat{H}^a_{(p+1)}=d\hat{B}_{(p)}^a$, $a=1,...,r$, transforming in
a real $r$-dimensional representation of the symmetry group
$\tilde{G}$. This example will play a central role in the coming
sections. The Lagrangian we will study is
\begin{equation}\label{actionhigh}
\mathcal{L} =  R \hat{*} 1 + \frac{1}{4} {\rm tr}(d \hat{\CK}
\wedge \hat{*} d\hat{\CK}^{-1})  - \frac{1}{2} \hat{H}_{(p+1)}^{t}
\hat{\CK}^{-1} \wedge \hat{*} \hat{H}_{(p+1)}.
\end{equation}
 Here $\hat{\CK}$ is an $r \times r$ matrix of scalar fields which act as a metric on the coset space
 $\tilde{G}/H$.  The
 Lagrangian (\ref{actionhigh}) is invariant under the rigid $\tilde{G}$ symmetry
\begin{equation}
\label{transform}  \hat{B}_{(p)} \to L \hat{B}_{(p)}, \qquad \
\hat{\CK}\to L \hat{\CK} L^t
\end{equation}
where $L ^a{}_b$ is a $\tilde{G}$-transformation in the {\bf r}
representation, and the space-time metric is invariant. The
invariance of the metric means that the ansatz for the dimensional
reduction of the metric is the same as the standard ansatz for the
group manifold reduction, which we presented in (\ref{gmfd}). On
the other hand,  the ansatz for the reduction of the scalar and
the p-form fields are dictated by their transformation
(\ref{transform}) under $\tilde{G}$, so that we have
\begin{eqnarray}
 \hat{\CK}(x, y)  =\tilde{g}(y) \CK(x) \tilde{g}^{t}(y), \label{ansatzscalar}
 \end{eqnarray}
 \begin{equation}
 \hat{B}_{(p)}(x, y)  =  \tilde{g}(y)\hat{B}_{(p)}(x), \label{ansatzA}
\end{equation}
where $\hat{B}_{(p)}(x)$ on the right hand side of (\ref{ansatzA})
is as in (\ref{pform}) and $\tilde{g}(y) \in \tilde{G}$. Note that
 $y^m$ here are coordinates on $G$, now that our internal space is
the group manifold $G$. In other words, $\tilde{g}$ is a map from
the group manifold to the duality group $\tilde{G}$: $\tilde{g}: G
\rightarrow \tilde{G}$. In the next subsection, we will see that
imposing the condition for $\tilde{g}^{-1}d\tilde{g}$ to be a
constant 1-form is necessary and also sufficient in order for the
dependence on the internal coordinates $y^m$ to cancel out in the
$D$ dimensional Lagrangian. Then in the following subsection, we
will see that a new condition has to be imposed for the lower
dimensional gauge algebra to close.

\subsection{Action}

In this section we impose the condition that
$\tilde{g}^{-1}d\tilde{g}$ has constant components with respect to
 the Maurer-Cartan forms of the group manifold $G$, so that it can be
 written as
\begin{equation}\label{onemli}
\tilde{g}^{-1}d\tilde{g} = M_n \eta^n,
\end{equation}
where $M_n$ are constant elements of the Lie algebra of
$\tilde{G}$.

The ansatz (\ref{ansatzA}) for the reduction of the p-form fields
$\hat{B}_{(p)}$ implies  for the field strength $\hat{H}_{(p+1)}$
\begin{eqnarray}\label{H}
\hat{H}_{(p+1)}(x^{\mu},y) &=& d\hat{B}_{(p)}(x^{\mu},y) =
d\tilde{g}(y) \hat{B}_{(p)}(x^{\mu}) + \tilde{g}(y)
d\hat{B}_{(p)}(x^{\mu}) \nonumber \\
&=& \tilde{g}(y)[d\hat{B}_{(p)}(x^{\mu}) +
\tilde{g}^{-1}d\tilde{g}(y) \hat{B}_{(p)}(x^{\mu})]
\end{eqnarray}
When we insert (\ref{H}) and (\ref{ansatzscalar}) in the kinetic
term for the p-form fields in (\ref{actionhigh}), we see that it
is independent of $y$ as the overall $\tilde{g}(y)$ factor cancels
out due to the global $\tilde{G}$ invariance of the action, and
$\tilde{g}^{-1}d\tilde{g}$ is independent of $y$ by
(\ref{onemli}). Now we check the kinetic term for the scalar
fields
\begin{eqnarray}\label{kinetics}
\CL_s &=& \frac{1}{4}{\rm tr}(d\hat{\CK} \wedge \hat{*}
d\hat{\CK}^{-1}).
\end{eqnarray}
From the Scherk-Schwarz ansatz (\ref{ansatzscalar}) for the scalar
fields we see
\begin{eqnarray}\label{yenibir}
d\hat{\CK} &=& d\tilde{g} \CK \tilde{g}^t + \tilde{g} d\CK
\tilde{g}^t + \tilde{g} \CK d\tilde{g}^t = \tilde{g}(M_p \CK
\eta^p + d\CK + \CK M_p^t \eta^p) \tilde{g}^t.
\end{eqnarray}
\noindent Similarly,
\begin{eqnarray}\label{yeniiki}
d\hat{\CK}^{-1} = (\tilde{g}^t)^{-1}(-M_p^t \CK^{-1} \eta^p +
d\CK^{-1} - \CK^{-1} M_p \eta^p)\tilde{g}^{-1},
\end{eqnarray}
\noindent where we have used $d(\tilde{g}^{-1})\tilde{g} = -
\tilde{g}^{-1} d\tilde{g} = - M_n \eta^n$. The overall $\tilde{g}$
factors cancel out in the action due to invariance of the
Lagrangian under (\ref{transform}). As the mass matrices $M_n$ are
constant, we conclude that the kinetic term for the scalar fields
is also independent of the internal coordinates.

\subsection{Gauge Algebra}
In this section we  choose $p=2$ for simplicity. This is also the
most interesting case for us, as we will see in section 5. In the
dimensional reduction of a $\tilde{G}$-invariant theory of the
form (\ref{actionhigh}), one obtains three sets of gauge fields in
the reduced theory. These are the fields $B^a_{(2)}$ and
$B^a_{(1)m}$ coming from the reduction of the 2-form fields
$B^a_{(2)}$ and the vector fields $\CA^m$ coming from the
reduction of the metric. Here $m$ runs from 1 to $d$, where $d$ is
the dimension of the internal twisted torus. We denote the
generators of the corresponding gauge transformations and the
gauge parameters $\{Y^a, X^{m a}, Z_m\}$ and $\{\Lambda^a_{(1)},
\Lambda^a_{(0)m}, \omega^m \}$, respectively. For a reduction on a
twisted torus, the gauge transformation for the vector fields are
always of the form \cite{SS2}
\begin{equation}\label{gaugeA}
\delta \CA^m = d\omega^m + C^m_{\ \ np} \omega^n \CA^p.
\end{equation}
Note that this implies that the $h^m = \eta^m + \CA^m$ always
transform covariantly
\begin{equation}\label{gaugeh}
\delta h^m =  C^m_{\ \ np} \omega^n h^p,
\end{equation}
as we have $i_{Z}\eta^m = -\omega^m$ and
\begin{equation}\label{gaugeeta}
\delta \eta^m = di_{Z} \eta^m + i_{Z} d\eta^m = -d\omega^m +
C^m_{\ \ np} \omega^n \eta^p.
\end{equation}
Before proceeding to compute the gauge transformations, let us
first write down the field strengths for gauge fields in lower
dimensions. From the ansatz for the 2-form field $B_2$
\begin{equation}\label{Bfield}
\hat{B}_{(2)}(x^{\mu},y^n) =\tilde{g}(y^n)[B_{(2)}(x^{\mu}) +
B_{(1)m}(x^{\mu}) \wedge h^m + \frac{1}{2!} B_{(0)mn}(x^{\mu}) h^m
\wedge h^n],
\end{equation}
 we find for the field strength $\hat{H}_{(3)} = d\hat{B}_{(2)}$
\begin{eqnarray}\label{H3}
\hat{H}_{(3)}(x^{\mu},y^n) &=& \tilde{g}(y^n)[H_{(3)}(x^{\mu}) +
H_{(2)n}(x^{\mu}) \wedge h^n \\
&&+ \frac{1}{2!}H_{(1)np}(x^{\mu}) \wedge h^n \wedge h^p +
\frac{1}{3!}H_{(0)mnp}(x^{\mu}) h^m \wedge h^n \wedge h^p]
\nonumber,
\end{eqnarray}
where
\begin{eqnarray}
\label{fieldstrength}
 && H_{(3)}  = dB_{(2)} - B_{(1)m} \wedge \CF^m - M_n B_{(2)} \wedge \CA^n \nonumber \\
  &&H_{(2)n} = dB_{(1)n} + B_{(1)m}C^m_{\ \ np} \CA^p + B_{(0)mn} \wedge \CF^m + M_n B_{(2)} + M_p B_{(1)n} \CA^p \nonumber \\
  &&H_{(1)np} =dB_{(0)np} + B_{(1)m}C^m_{\ \ np} + B_{(0)m[n}C^m_{\ \ p]r} \CA^r - M_{[n}B_{(1)p]} - M_r B_{(0)np} \CA^r \nonumber   \\
 && H_{(0)npr} = M_{[n}B_{(0)pr]} + C^m_{\ [np}B_{(0)r]m}
\end{eqnarray}
Here
\begin{equation}\label{F}
\CF^m = d \CA^m - \frac{1}{2} C^m_{ \ \ np} \CA^n \wedge \CA^p,
\end{equation}
and we have used
\begin{eqnarray}\label{faydali}
dh^m &=& d \eta^m + d \CA^m = \CF^m + C^m_{\ \ np} \CA^n \wedge
h^p - \frac{1}{2} C^m_{\ \ np} h^n \wedge h^p.
\end{eqnarray}
Note that $\CF^m$ varies covariantly
\begin{equation}\label{gaugeF}
\delta \CF^m = C^m_{\ \ np} \omega^n \CF^p.
\end{equation}
The terms in (\ref{fieldstrength}) involving the structure
constants $C^m_{\ \ np}$ are due to the twisted internal geometry,
whereas the terms involving the mass matrices $M_p$ are due to the
twisted ansatz for the 2-form field $B_{(2)}$.

Next we find the gauge transformations of the 2-form, 1-form and
the scalar fields coming from the reduction of $\hat{B}_{(2)}$
inherited from its diffeomorphism invariance and the gauge
transformation $\delta \hat{B}_{(2)} =
d\hat{\Lambda}_{(1)}(x^{\mu},y)$. Let us start with the latter. As
we are using a generalized ansatz (\ref{Bfield}) for the reduction
of $\hat{B}_{(2)}$, the ansatz for the higher dimensional gauge
parameter $\hat{\Lambda}_{(1)}(x^{\mu},y^m)$ should be
\begin{equation}\label{ansatzlambda}
\hat{\Lambda}_{(1)}(x^{\mu},y^m) =
\tilde{g}(y^m)[\Lambda_{(1)}(x^{\mu}) + \Lambda_{(0)m}(x^{\mu})
\wedge h^m].
\end{equation}
From (\ref{ansatzlambda}) we find
\begin{eqnarray}\label{dlambda}
d\hat{\Lambda}_{(1)} &=& \tilde{g}(y)[d\Lambda_{(1)} +
\Lambda_{(0)m}\wedge \CF^m + M_p \Lambda_{(1)} \CA^p \nonumber \\
&&+(d\Lambda_{(0)m} + \Lambda_{(0)n}C^n_{\ \ pm} \CA^p - M_m
\Lambda_{(1)} - M_p \Lambda_{(0)m} \CA^p) \wedge h^m \nonumber \\
&&+\frac{1}{2!}(-\Lambda_{(0)p}C^p_{\ \ mn} + M_{[m}
\Lambda_{(0)n]})h^m \wedge h^n]
\end{eqnarray}
Comparing this to $\delta \hat{B}_{(2)}$ with the ansatz
(\ref{Bfield})  we find
\begin{eqnarray}\label{bir}
& & \delta B_{(2)}  =  d\Lambda_{(1)} + M_{p } \Lambda_{(1)} \CA^p + \Lambda_{(0)m} \CF^m \\
& & \delta B_{(1)m} = d\Lambda_{(0)m} - \Lambda_{(0)n} C^n_{\ \
mp} \CA^p - M_{m } \Lambda_{(1)} -
M_{p } \Lambda_{(0)m} \CA^p \nonumber \\
& & \delta B_{(0)mn}  =  -\Lambda_{(0)p} C^p_{\ \ mn} + M_{[m }
\Lambda_{(0)n]} \nonumber
\end{eqnarray}
Now we find the gauge transformations inherited from the
diffeomorphism invariance of $\hat{B}_{(2)}$.
\begin{eqnarray}
\delta_{Z}\hat{B}_{(2)} &=& \CL_{Z}\hat{B}_{(2)} = 0 =
\CL_{Z}[\tilde{g}(y)(B_{(2)} + B_{(1)m} \wedge h^m + \frac{1}{2!}
B_{(0)mn} h^m \wedge h^n)] \nonumber \\
&=& (\CL_{Z} \tilde{g}(y)) (B_{(2)} + B_{(1)m} \wedge h^m +
\frac{1}{2!}
B_{(0)mn} h^m \wedge h^n) \nonumber \\
&&+ \tilde{g}(y) \CL_{Z}(B_{(2)} + B_{(1)m} \wedge h^m +
\frac{1}{2!} B_{(0)mn} h^m \wedge h^n).
\end{eqnarray}
Noting $$\CL_{Z} \tilde{g}(y) = i_{Z} d\tilde{g}(y) + d i_{Z}
\tilde{g}(y) =i_{Z}
d\tilde{g}(y)=\tilde{g}i_{Z}\tilde{g}^{-1}d\tilde{g}(y) = -
\tilde{g} M_p \omega^p,$$  and taking into account (\ref{gaugeh})
we obtain
\begin{eqnarray}\label{iki}
& & \delta_Z B_{(2)}  =  M_p \omega^p B_{(2)} \\
& & \delta_Z B_{(1)m}  =  M_p \omega^p B_{(1)m} - B_{(1)p}
C^p_{\ \ nm} \omega^n \nonumber \\
& & \delta_Z B_{(0)mn}  =  M_p \omega^p B_{(0)mn} + B_{(0)p[n}
C^p_{\ \ m]q} \omega^q \nonumber
\end{eqnarray}
One can check that the field strengths (\ref{fieldstrength}) are
indeed invariant under the gauge transformations (\ref{bir}) and
transform covariantly under (\ref{iki}).

Now we have to check that the gauge algebra closes. At this point,
we find that we need to impose a new condition in addition to the
constancy of the mass matrices $M_n$. Namely, $M_n$ has to satisfy
the following commutation relations
\begin{equation}\label{comm}
[M_n, M_p] = C^{q}_{\ \ np} M_q.
\end{equation}
This means that $M_n$, which are elements of the Lie algebra of
$\tilde{G}$ should form a representation of the Lie algebra of
$G$. Obviously, this is possible only if $G$ is a subgroup of
$\tilde{G}$. Note that we should have anticipated  the condition
(\ref{comm}) already in the previous subsection, when we imposed
that the mass matrices $M_n$ in (\ref{onemli}) do not depend on
the coordinates $y^m$ of $G$. To see why, we should first note
that the 1-form $A \equiv \tilde{g}^{-1}d\tilde{g}$ automatically
satisfies the zero curvature condition
\begin{equation}\label{connection}
dA + A \wedge A = 0.
\end{equation}
On the other hand, the basis 1-forms $\eta^m$ satisfy the
Maurer-Cartan equation  (\ref{structure}). These two equations are
compatible with constant $M_p$ if and only if the commutation
relation (\ref{comm}) is satisfied. Therefore, our analysis of the
gauge algebra shows that it is consistent to require $A$ to be a
constant, flat $\tilde{G}$-connection on $G$. At this point, we
see that the topology of $G$ plays an important role in analyzing
as to whether it is possible to introduce non-trivial duality
twists on the group manifold $G$. More precisely, we see that if
$\pi_1(G)=0$ so that the group manifold is simply-connected, then
the flat connection $A$ is pure gauge globally, giving a
dimensional reduction equivalent to a standard group manifold
reduction with no duality twist. This follows from the fact that
the moduli space of flat $\tilde{G}$-connections on $G$ (modulo
smooth gauge transformations) can be identified with
$Hom(\pi_1(G),\tilde{G})/\tilde{G}$, where $\tilde{G}$ acts by
conjugation. Therefore, it is crucial that $G$ is
non-simply-connected so that the non-trivial Wilson
lines/holonomies of the connection $A$ over the cycles of $G$
introduce twisted boundary conditions for the fields charged under
$\tilde{G}$ (with analogy to a twisted reduction on $T^d$ in which
case we have the twisted boundary conditions (\ref{bc})). So far
our analysis has been only local and we will not be studying the
global issues in the rest of the paper, either. However, we will
have a bit more to say on the non-simply-connectedness condition
on the internal space, when we restrict ourselves to particular
examples in the next section.

When  the condition (\ref{comm}) is satisfied, the gauge algebra
has the following form\footnote{In finding the gauge algebra, we
also have to use the condition $C^m_{\ [np}C^q_{\ r]m}=0$, which
follows immediately from the integrability of (\ref{structure}).}
\begin{equation}\label{algebra1}
[Z_n , X^{m}] = C^m_{\ \ nq} X^{q} - M_{n } X^{m}
\end{equation}
\begin{equation}\label{algebra2}
[ Z_{p} , Y^{a} ]  =  -M_{p \ b}^a Y^{b}
\end{equation}
\begin{equation}\label{algebra3}
 [Z_p , Z_q]  = -C^r_{\ \ pq} Z_r.
\end{equation}
Note that this algebra is different from the one that is obtained
from the reduction on a twisted torus with constant flux for the
2-form field $B^{a}_{(2)}$. In this case, it was shown in
\cite{Dall'Agata:2005ff,Hull:2005hk} that the gauge
transformations coming from the higher dimensional gauge
invariance of the 2-form field is still of the form (\ref{bir}),
where now all the terms with the mass matrices $M_p$ are zero. On
the other hand the gauge transformations coming from the
diffeomorphism invariance of $B_{(2)}$ is of the form
\begin{eqnarray}\label{flux}
& & \delta_Z B_{(2)}  =  \frac{1}{2} K_{mnp} \omega^p \CA^m \wedge \CA^n  \\
& & \delta_Z B_{(1)m}  =  K_{mnp}\omega^p \CA^n - B_{(1)p}
C^p_{\ \ nm} \omega^n \nonumber \\
& & \delta_Z B_{(0)mn}  =  K_{mnp} \omega^p + B_{(0)p[n} C^p_{\
m]q} \omega^q, \nonumber
\end{eqnarray}
where
$$\frac{1}{3!} K_{mnp} \eta^m \wedge \eta^n \wedge \eta^p$$
with constant $K_{mnp}$ being  the 3-form flux introduced for the
2-form field $B_{(2)}$. We see that (\ref{flux}) is equivalent to
(\ref{iki}) only for special values of the gauge fields, namely
when
\begin{eqnarray}\label{on}
&& M_{p } B_{(2)} = \frac{1}{2} K_{pmn} \CA^m \wedge
\CA^n \nonumber \\
&& M_{[p} B_{(1)m]} = - K_{pmn} \CA^n \nonumber \\
&& M_{[p } B_{(0)mn]} = - K_{pmn}
\end{eqnarray}

As a final remark, note that a gauge algebra of the form
(\ref{algebra1},\ref{algebra2},\ref{algebra3}) was obtained in
\cite{myers1}, albeit in a completely different context. There
they consider the compactifications of the heterotic string, the
generators $Y^{a}$ are associated with the gauge invariance of the
16 vector fields in the Yang-Mills sector that already exist in 10
dimensions  and $M_p$ correspond to the internal fluxes in the
Yang-Mills sector.


\section{Three Dimensional Group Manifolds and Twisted Tori}
In this section we restrict ourselves to three dimensional group
manifolds. The classification of three dimensional algebras was
made long ago by Bianchi. As reviewed by \cite{Roest:2004pk} there
are eleven inequivalent three dimensional algebras, two of which
are one-parameter families. Among them, five are of Type B,
meaning that they are not unimodular algebras. The six unimodular
algebras of Type A include the two semi-simple algebras $so(2,1)$
and $so(3)$. We are interested here in the three non-abelian
unimodular non-semisimple algebras: $heis_3$, $iso(1,1)$ and
$iso(2)$. The fourth unimodular, non-semi-simple algebra, which is
the only abelian three dimensional algebra, $u(1)^3$ has the group
manifold $T^3$ (after some discrete identifications, which we are
not interested here as we study only the local structure), so the
corresponding group manifold reduction is a standard Kaluza-Klein
reduction on a three dimensional torus. As we mentioned before, we
require unimodularity  so that there is a consistent dimensional
reduction at the level of the action. On the other hand,
non-semi-simplicity is required because then the corresponding
group manifold has the local structure of a $T^2$ fibration over
$S^1$ \cite{Bergshoeff:2003ri},\cite{Hull:2005hk}, making it
easier to analyze the twisted torus geometry. We will now study
this structure, following closely the discussion in
\cite{Hull:2005hk}.

Consider a reduction on $T^2$ with metric
\begin{equation}\label{metrictorus1}
ds_2^2=\frac{A}{\tau_2}\mid \tau dx_1 + dx_2 \mid^2
\end{equation}
where $\tau = \tau_1 + i \tau_2$ is the complex structure modulus
and $A$ is the area of $T^2$. The metric (\ref{metrictorus1}) can
also be written as
\begin{equation}\label{metrictorus2}
ds_2^2=H(\tau)_{ab}dx^a dx^b = dx^t H(\tau) dx,
\end{equation}
with
\begin{equation}\label{HH}
H(\tau) = \frac{A}{\tau_2}\left(\begin{array}{cc}
                                \mid \tau \mid^2 & \tau_1 \\
                                \tau_1 & 1
                                \end{array}\right).
\end{equation}
The symmetry group associated with the large diffeomorphisms of
$T^2$ is $GL(2,\IR)$, with the volume preserving subgroup
$SL(2,\IR)$. The action of $SL(2,\IR)$ on the metric is
\begin{equation}\label{tr1}
H \rightarrow L H L^t, \ \ \ \ \ x \rightarrow (L^t)^{-1} x
\end{equation}
with $L \in SL(2,\IR)$. This defines the transformation of the
moduli through
\begin{equation}\label{tr2}
H(\tau') = L H(\tau) L^t.
\end{equation}
If $$L = \left(\begin{array}{cc}
                        a & b \\
                        c & d
                        \end{array}\right) \in SL(2,\IR)$$ then
(\ref{tr2}) is equivalent to
\begin{eqnarray}\label{tr3}
\tau_1 & \longrightarrow & \frac{ac(\tau_1^2 + \tau_2^2) +
(ad+bc)\tau_1 + bd}{c^2 (\tau_1^2 + \tau_2^2) + 2dc \tau_1 + d^2},
\nonumber \\
\tau_2 & \longrightarrow & \frac{\tau_2}{c^2 (\tau_1^2 + \tau_2^2)
+ 2dc \tau_1 + d^2}.
\end{eqnarray}
The transformation (\ref{tr3}) of the moduli can be written in the
compact form
\begin{equation}\label{tr4}
\tau \rightarrow \frac{a \tau + b}{c \tau + d} \equiv L[\tau],
\end{equation}
which of course is nothing but the fractional linear
transformation of the complex structure modulus of $T^2$ under the
action of $SL(2,\IR)$.

After the dimensional reduction on $T^2$ the geometric internal
symmetry $SL(2,\IR)$ of $T^2$ is promoted to become an external
symmetry of the lower dimensional theory, which we can use to
perform a further reduction on a circle with duality twist. One
can introduce a twisted ansatz for the two massless scalar fields
corresponding to the moduli $\tau_i$ through
\begin{equation}\label{ansatzh}
H(\tau(y)) = s(y) H(\tau_0) s^t(y)
\end{equation}
with $s(y)$ a $y$ dependent $SL(2,\IR)$ element and $\tau_0$ is a
constant value of the modulus. Here $y$ parameterizes the circle
$S^1$. The ansatz (\ref{ansatzh}) follows directly from the
transformation (\ref{tr2}) of $\tau$. Then the metric of the three
dimensional total space is
\begin{equation}\label{metrictorus3}
ds_3^2  = dy^2 +  dx^t H(\tau(y)) dx = (\eta^y)^2 + \eta^t
H(\tau_0) \eta,
\end{equation}
where
\begin{equation}\label{eta}
\eta^y = dy, \ \ \  \eta^a(y) = (s^t(y))^a_{ \ \ b} dx^b,
\end{equation}
with $a,b = 1,2$. The metric (\ref{metrictorus3}) is equivalent to
(\ref{torusmetric}).

The group structure of this space was studied in
\cite{Hull:2005hk}. The globally well-defined 1-forms (\ref{eta})
satisfy
\begin{equation}
d\eta^a + (N^t)^a_{\ b} \eta^y \wedge \eta^b = 0,
\end{equation}
where  $N$ is the Lie algebra element $s(y) =
e^{Ny}$. 
 Then locally, the space with the metric
(\ref{metrictorus3}) has the structure of a group manifold with
Maurer-Cartan 1-forms (\ref{eta}). The associated Lie algebra is
\begin{equation}\label{alg}
[t_a, t_y] = N_a^{\ b} t_b, \ \ \ \ [t_a, t_b]=0.
\end{equation}

It is well known that reductions with duality twists are
classified according to the conjugacy classes of the duality
group. The symmetry group $SL(2,\IR)$ of our interest here has
three conjugacy classes: parabolic, elliptic and hyperpolic.

~

\noindent \textit{Parabolic Conjugacy Class:}

In this case the group element $s(y)$ and the corresponding matrix
$N$ are
\begin{equation}\label{para}
s(y) = \left(\begin{array}{cc}
                   1 & my \\
                   0 & 1
                   \end{array}\right), \ \ \ \ \ N =
                   \left(\begin{array}{cc}
                                   0 & m \\
                                   0 & 0
                                   \end{array}\right).
\end{equation}
Then the Lie algebra (\ref{alg}) is the Heisenberg algebra
$heis_3$, also called the Bianchi II in the Bianchi classification
scheme, which is a non-semi-simple algebra of Type A. The
Maurer-Cartan 1-forms (\ref{eta}) are
\begin{equation}\label{eta2}
\eta^1 = dx^1, \ \ \ \eta^2 = dx^2 + m y dx^1, \ \ \ \ \eta^3 =
dy.
\end{equation}
In summary, reducing on $T^2$ followed by a circle reduction with
a duality twist in the parabolic conjugacy class of $SL(2,\IR)$ is
locally equivalent to reducing from $D+3$ to $D$ dimensions on the
group manifold of the Heisenberg algebra (also known as the
nilmanifold) with the Maurer-Cartan 1-forms (\ref{eta2}). Now
suppose that the $D+3$ dimensional theory that we start with has a
duality symmetry $\tilde{G}$ which contains the Heisenberg group
as a subgroup. Then, as we saw in the previous section, it is
possible to introduce a twisted ansatz for the dimensional
reduction of the $D+3$ fields on the nilmanifold through a group
element $\tilde{g}(x^1,x^2,y)$ provided that it satisfies
(\ref{onemli}). For consistency, the mass matrices $M_n$ must be
constant and form a representation of the Heisenberg algebra. If
$\tilde{G}$ is the Heisenberg group itself, then a convenient
choice of $\tilde{g}$ is
\begin{equation}\label{twistpara}
\tilde{g} = \left(\begin{array}{ccc}
                         1 & -q y & p x^2 \\
                         0 & 1 & r x^1 \\
                         0 & 0 & 1
                         \end{array}\right).
\end{equation}
Then the mass matrices are
\begin{equation}\label{twistparamass}
 M_1 = \left(\begin{array}{ccc}
                 0 & 0 & 0 \\
                 0 & 0 & r \\
                 0 & 0 & 0
                 \end{array}\right), \ \ \ \ M_2 = \left(\begin{array}{ccc}
                 0 & 0 & p \\
                 0 & 0 & 0 \\
                 0 & 0 & 0
                 \end{array}\right), \ \ \ \ M_3 = \left(\begin{array}{ccc}
                 0 & -q & 0 \\
                 0 & 0 & 0 \\
                 0 & 0 & 0
                 \end{array}\right).
\end{equation}
Here $p, q, r$ have dimension of mass and satisfy $\frac{qr}{p} =
m$. One can check that $M_n$ form a representation of the
Heisenberg algebra and that $\tilde{g}$ in (\ref{twistpara})
satisfies (\ref{onemli}) with $M_n$ in (\ref{twistparamass}) and
$\eta^m$ in (\ref{eta2}).

~

\noindent \textit{Elliptic Conjugacy Class:}

In this case, the group element $s(y)$ and the corresponding
matrix $N$ are
\begin{equation}\label{elliptic}
s(y) = \left(\begin{array}{cc}
                   \cos{my} & \sin{my} \\
                   -\sin{my} & \cos{my}
                   \end{array}\right), \ \ \ \ \ N =
                   \left(\begin{array}{cc}
                                   0 & m \\
                                   -m & 0
                                   \end{array}\right).
\end{equation}
The basis 1-forms (\ref{eta}) are
\begin{equation}\label{eta3}
\eta^1 = \cos{my} dx^1 - \sin{my} dx^2, \ \ \ \ \eta^2 =
\sin{my}dx^1 + \cos{my} dx^2, \ \ \ \ \eta^3 = dy.
\end{equation}
The corresponding algebra (\ref{alg}) is that of $iso(2)$ (Bianchi
VII$_0$), so we see that a $T^2$ reduction followed by a reduction
on a circle with an $SL(2,\IR)$ twist in the elliptic conjugacy
class is locally equivalent to group manifold reduction on
$ISO(2)$. If we start with a $D+3$ dimensional theory invariant
under a symmetry group $\tilde{G}$ that contains $ISO(2)$, then we
can introduce a duality twist for the reduction of the
$D+3$-dimensional fields through a group element $\tilde{g} \in
\tilde{G}$ provided that $\tilde{g}$ satisfies (\ref{onemli}) with
constant $M_n$ forming a representation of $iso(2)$. If
$\tilde{G}$ is $ISO(2)$ itself, a convenient choice of $\tilde{g}$
is
\begin{equation}\label{twistelliptic}
\tilde{g} = \left(\begin{array}{ccc}
                         \cos{my} & \sin{my} & p x^1 \\
                         -\sin{my} & \cos{my} & p x^2 \\
                         0 & 0 & 1
                         \end{array}\right).
\end{equation}
Then the mass matrices are
\begin{equation}\label{twistellipticmass}
M_1 = \left(\begin{array}{ccc}
                 0 & 0 & p \\
                 0 & 0 & 0 \\
                 0 & 0 & 0
                 \end{array}\right), \ \ \ \ M_2 = \left(\begin{array}{ccc}
                 0 & 0 & 0 \\
                 0 & 0 & p \\
                 0 & 0 & 0
                 \end{array}\right), \ \ \ \ M_3 = \left(\begin{array}{ccc}
                 0 & m & 0 \\
                 -m & 0 & 0 \\
                 0 & 0 & 0
                 \end{array}\right).
\end{equation}

~

\noindent \textit{Hyperbolic Conjugacy Class:}

In this case, the group element $s(y)$ and the corresponding
matrix $N$ are
\begin{equation}\label{elliptic}
s(y) = \left(\begin{array}{cc}
                   e^{my} & 0 \\
                   0 & e^{-my}
                   \end{array}\right), \ \ \ \ \ N =
                   \left(\begin{array}{cc}
                                   m & 0 \\
                                   0 & -m
                                   \end{array}\right).
\end{equation}
The basis 1-forms (\ref{eta}) are
\begin{equation}\label{eta4}
\eta^1 = e^{my} dx^1 \ \ \ \ \eta^2 = e^{-my} dx^2, \ \ \ \ \eta^3
= dy.
\end{equation}
The corresponding algebra (\ref{alg}) is that of $iso(1,1)$
(Bianchi VI$_0$), so we see that a $T^2$ reduction followed by a
reduction on a circle with an $SL(2,\IR)$ twist in the hyperbolic
conjugacy class is locally equivalent to group manifold reduction
on $ISO(1,1)$. If  the $D+3$ dimensional theory we start with is
invariant under a symmetry group $\tilde{G}$ that contains
$ISO(1,1)$, then we can introduce a duality twist for the
reduction of the $D+3$-dimensional fields through a group element
$\tilde{g} \in \tilde{G}$ provided that $\tilde{g}$ satisfies
(\ref{onemli}) with constant $M_n$ forming a representation of
$iso(1,1)$. If $\tilde{G}$ is $ISO(1,1)$ itself, a convenient
choice of $\tilde{g}$ is
\begin{equation}\label{twisthyper}
\tilde{g} = \left(\begin{array}{ccc}
                         e^{-my} & 0 & p x^1 \\
                         0 & e^{my} & q x^2 \\
                         0 & 0 & 1
                         \end{array}\right).
\end{equation}
Then the mass matrices are
\begin{equation}\label{twisthypermass}
M_1 = \left(\begin{array}{ccc}
                 0 & 0 & p \\
                 0 & 0 & 0 \\
                 0 & 0 & 0
                 \end{array}\right), \ \ \ \ M_2 = \left(\begin{array}{ccc}
                 0 & 0 & 0 \\
                 0 & 0 & q \\
                 0 & 0 & 0
                 \end{array}\right), \ \ \ \ M_3 = \left(\begin{array}{ccc}
                 -m & 0 & 0 \\
                 0 & m & 0 \\
                 0 & 0 & 0
                 \end{array}\right).
\end{equation}
Note that we can introduce three independent mass parameters in
this case, whereas it was only possible to introduce  two
independent mass parameters in the previous cases.

Before we close this section, we would like to make a remark on
the global issues, or more precisely on the condition of
non-simply-connectedness, which comes about from requiring that
the connection $A = \tilde{g}^{-1}d\tilde{g}$ is not pure gauge
globally. The requirement that the internal space has a
non-trivial fundamental group is clearly a topological condition,
so it cannot be explored by the local analysis we have pursued
here. For example, corresponding to the  Bianchi IX algebra
$so(3)$, there are two groups: $SO(3)$ and $SU(2)$. The latter is
a simply-connected group with trivial fundamental group, whereas
$SO(3)$ has $\pi_1(SO(3))=Z_2$. A more detailed analysis than we
have carried out here, including the global issues was given in
\cite{Hull:2005hk}. Note that all three group manifolds that we
have studied in this section are non-compact. Performing a
dimensional reduction on a non-compact group manifold $G$ leads to
a continuous spectrum in the lower dimensional theory. However, it
is possible to consistently truncate this spectrum to a finite
number of fields, yielding gauged supergravities if the higher
dimensional theory that we start with is itself a supergravity
theory. This can be extended to a compactification of string
theory only if one can construct a compact internal space $X =
G/\Gamma$ by compactifying $G$ by dividing out by the action of a
discrete symmetry group $\Gamma \subset G$ \cite{Hull:2005hk}. For
all the Bianchi types except type IV and VI$_a$ it is possible to
construct compact manifolds in this way
\cite{Roest:2004pk}\footnote{See \cite{Hull:2005hk}  for the
explicit forms of the discrete subgroups $\Gamma$ which one can
use to compactify the three non-abelian, unimodular,
non-semi-simple group manifolds we have considered here.}.
Therefore, even if the non-compact group manifold that we start
with is simply-connected so that $\pi_1(G)$ is trivial, the
compact manifold $X = G/\Gamma$ that we construct from $G$ will
have a non-trivial fundamental group $\pi_1(X) \approx \Gamma$.
This then makes it possible to introduce non-trivial duality
twists on $X$, as we discussed in section 3.


\section{Example}

In this section we will study the dimensional reduction of a
particular eight dimensional theory with a duality twist on a
three dimensional unimodular, non-semi-simple group manifold. As
we saw in the previous section, there are three such group
manifolds (other than the abelian $T^3$) and locally each has the
structure of a twisted torus. The eight dimensional theory that we
are interested in is the Type II supergravity, which has the
duality symmetry $SL(2,\IR) \times SL(3,\IR)$ \cite{Hull:1994ys}.
This theory describes the low energy effective theory of Type
IIA/IIB string theory compactified on $T^2$, or equivalently
M-theory compactified on $T^3$ \cite{salam,Alonso-Alberca:2000gh}.
It has a consistent truncation to the sector of $SL(2,\IR)$
singlets, which is described by the Lagrangian\footnote{Here we
are omitting the hats on the higher dimensional fields, hoping
that the eight-dimensional fields here will not be confused with
the five-dimensional fields in (\ref{scalar5d},\ref{kinetic5d})
and (\ref{cs5d}).}
\begin{equation}\label{8d}
\CL_8 = R * 1 + \frac{1}{4} {\rm Tr}(d\CK \wedge * d\CK^{-1}) +
\frac{1}{2} H_{(3)}^t\CK^{-1} \wedge * H_{(3)} + \frac{1}{6}
\epsilon^{abc} H_{(3)a} \wedge H_{(3)b} \wedge B_{(2)c}
\end{equation}
where $a,b,c=1,2,3$ and $H_{(3)}=dB_{(2)}$ is the 3-vector
$$H_{(3)} = \left(\begin{array}{c}
                        dB_{(2)1} \\
                        dB_{(2)2} \\
                        dB_{(2)3}
                        \end{array}
                       \right).$$
From the M-theory point of view, the three 2-form fields
$B_{(2)a}$ come from the reduction of the eleven dimensional
3-form field. The IIB origin of $B_{(2)a}$ is the two (one in the
NS-NS sector and one in the R-R sector) 2-form fields that already
exist in the massless spectrum of the ten dimensional IIB string
theory and the self-dual 4-form field $C_{(4)}$. The dimensional
reduction of $C_{(4)}$ yields in eight dimensions one 2-form field
and one 4-form field, which can be dualized to a second 2-form
field. Imposing the self-duality constraint reduces the number of
2-form fields to one. The scalar matrix $\CK$ in (\ref{8d})
represents the coset space $SL(3,\IR)/SO(3)$ parameterized by the
scalar fields coming from the reduction of the eleven dimensional
metric. From the IIB point of view, two of the five scalars
parameterizing $\CK$ are the axion and the dilaton that exist in
ten dimensions; two come from the reduction of the two ten
dimensional 2-form fields and the fifth is the massless scalar
associated with the volume modulus of $T^2$ (the scalar coming
from the reduction of the 2-form field in the NS-NS sector
combines with this fifth scalar to form the K\"{a}hler modulus of
$T^2$). The Lagrangian (\ref{8d}) is manifestly invariant under
the duality group $SL(3,\IR)$ under which $B_{(2)}$ and $\CK$
transform as
\begin{equation}\label{tr8d}
B_{(2)} \rightarrow \Gamma B_{(2)}, \ \ \ \ \CK \rightarrow \Gamma
\CK \Gamma^t, \ \ \ \ \ \Gamma \in SL(3,\IR).
\end{equation}
One can use this $SL(3,\IR)$ symmetry to introduce a twisted
ansatz for the dimensional reduction of the fields $B_{(2)}$ and
$\CK$ through
\begin{equation}\label{ansatz8dB}
B_{(2)}(x^{\mu},y^m) = \tilde{g}(y^m) B_{(2)}(x^{\mu}) \nonumber
\end{equation}
\begin{equation}\label{ansatz8dM}
\CK(x^{\mu},y^m) = \tilde{g}(y^m) \CK(x^{\mu}) \tilde{g}^t(y^m).
\end{equation}
Here $y^m$ are the coordinates of the internal space and
$\tilde{g}(y^m) \in \tilde{G}=SL(3,\IR)$. As we mentioned above,
we would like to choose the internal space as a three dimensional
unimodular, non-semi-simple group manifold. We have seen in
section 3 that this group manifold reduction with the twisted
ansatz (\ref{ansatz8dB},\ref{ansatz8dM}) is consistent only if the
$sl(3,\IR)$-valued 1-form $A \equiv \tilde{g}^{-1}d\tilde{g}$ is a
constant 1-form $A = M_p \eta^p$ with the mass matrices $M_p$
satisfying the commutation relations (\ref{comm}). One can easily
find such $\tilde{g} \in SL(3,\IR)$. In fact, the group elements
we introduced in (\ref{twistpara}), (\ref{twistelliptic}) and
(\ref{twisthyper}) are all elements of the  group $SL(3,\IR)$.
Therefore, it is possible to introduce non-trivial $SL(3,\IR)$
twists through these choices of $\tilde{g}$ for the dimensional
reduction on the corresponding group manifolds. This gives a
gauged supergravity theory in five dimensions, with  the mass
matrices  determining the gauge and mass parameters and the scalar
potential given by (\ref{twistparamass}),(\ref{twistellipticmass})
and (\ref{twisthypermass}) for the nilmanifold, $ISO(2)$ and
$ISO(1,1)$ reductions respectively. Below we present the
Lagrangian describing the resulting five dimensional gauged
supergravity theory, leaving the details of the reduction to
Appendix A.

The five dimensional Lagrangian is
\begin{equation}\label{lag5d}
\CL_5 = \CL_{EH} + \CL_S + \CL_K + \CL_{CS}.
\end{equation}
Here $\CL_{EH}$ comes from the reduction of the Einstein-Hilbert
term in (\ref{8d}) and is exactly of the form (\ref{EHgmfd}) as we
are reducing on a group manifold. The eight dimensional metric is
invariant under the duality symmetry $SL(3,\IR)$, so there is no
correction to (\ref{EHgmfd}) due to the $SL(3,\IR)$ twist. The
 $\CL_S$ term in (\ref{lag5d}) comes from the reduction of the scalars parameterizing
$\CK$ and is of the form
\begin{equation}\label{scalar5d}
\CL_S = \frac{1}{4}{\rm Tr}(D\CK \wedge * D\CK^{-1}) -
e^{4\alpha\phi} V,
\end{equation}
where
\begin{eqnarray}\label{yeniuc}
D\CK &=& d\CK - M_m \CK \CA^m - \CK M^t_m \CA^m \nonumber \\
D\CK^{-1} &=& d\CK^{-1} + M_m^t \CK^{-1} \CA^m + \CK^{-1} M_m
\CA^m
\end{eqnarray}
and
\begin{equation}\label{potential5d}
V = \frac{1}{2} \sum_i {\rm Tr}(M_i^2 + \CK M_i^t \CK^{-1} M_i).
\end{equation}
Here $M_i = L_i^{\ m} M_m$ with $L_i^{\ m} L_j^{\ n} \delta^{ij} =
\CM^{mn}$. The $m,n$ are curved indices for the internal space,
whereas $i,j$ are the flat indices of the tangent space of the
internal space. (See Appendix A for details.) $M_m$ are the mass
matrices in (\ref{twistparamass}),(\ref{twistellipticmass}) or
(\ref{twisthypermass}) depending on which unimodular,
non-semisimple group manifold  we pick to reduce on and $\CM$ is
the matrix in the coset space $SL(3,\IR)/SO(3)$ parameterized by
the five scalars coming from the reduction of the eight
dimensional metric on the group manifold.

The dimensional reduction of the kinetic term for the fields
$B_{(2)}$ in (\ref{8d}) yields in five dimensions the following
kinetic terms
\begin{eqnarray}\label{kinetic5d}
\CL_K &=& \frac{1}{2}(e^{-4\alpha\phi}H_{(3)a}\CK^{ab}\wedge *
H_{(3)b} +
H_{(2)am}\CK^{ab}\CM^{mn} \wedge * H_{(2)bn} \nonumber \\
& & + e^{2\alpha\phi}H^m_{(1)a}\CK^{ab}\CM_{mn} \wedge *
H^n_{(1)b} + e^{8\alpha\phi} H_{(0)a}\CK^{ab} \wedge * H_{(0)b}).
\end{eqnarray}
Here  $H^m_{(1)} = \epsilon^{mnp} H_{(1)np}, \ H_{(0)} =
\epsilon^{mnp} H_{(0)mnp}$ and $H_{(3)}, H_{(2)m}, H_{(1)np},
H_{(0)mnp}$ are as in (\ref{fieldstrength}). Finally, the
dimensional reduction of the topological term in (\ref{8d}) gives
the $\CL_{CS}$ term in (\ref{lag5d})
\begin{equation}\label{cs5d}
\CL_{CS} = \epsilon^{abc}\epsilon^{mnp}(H_{(2)am} \wedge H_{(2)bn}
\wedge B_{(1)cp} + H_{(3)a} \wedge H_{(1)bmn} \wedge B_{(1)cp}).
\end{equation}
Note that there are two potential terms in (\ref{lag5d}). The
first is for the scalar fields coming from the reduction of the
eight  dimensional metric, associated with the moduli
parameterizing the internal space and is of the form
(\ref{potential}). The second term (\ref{potential5d}) is for
these geometric moduli and for the scalars that already exist in
eight dimensions, parameterizing the matrix $\CK$. As we mentioned
above, from the point of view of M-theory, these are geometric
moduli associated with the internal $T^3$ on which M-theory has
been compactified to obtain the eight dimensional theory
(\ref{8d}). On the other hand, from the point of view of IIB
theory, this is a potential term for the axion-dilaton field, the
K\"{a}hler modulus of the internal $T^2$ (on which IIB has been
compactified) and the scalar field coming from the reduction of
ten dimensional R-R 2-form field. The gauge algebra of our new
five dimensional gauged supergravity theory (\ref{lag5d}) is
exactly the algebra that we presented in
(\ref{algebra1},\ref{algebra2},\ref{algebra3}).


\section{Outlook}
In this paper we have studied the consistency conditions for the
dimensional reduction of a $\tilde{G}$-invariant theory with a
duality twist in $\tilde{G}$ on a group manifold $G$. We have seen
that if the duality twist is introduced through a group element
$\tilde{g} \in \tilde{G}$, then the $\tilde{G}$-connection $A =
\tilde{g}^{-1}d\tilde{g}$ on $G$ must be constant. So, if $\eta^m$
are basis 1-forms for the group manifold $G$ satisfying
(\ref{structure}) then $A$ can be written as $A = M_n \eta^n$ with
constant $M_n$ satisfying (\ref{comm}). $M_n$ are elements of the
Lie algebra of $\tilde{G}$ and introduce the mass and gauge
parameters in the lower dimensional theory.

Duality symmetries arise naturally in string theory and in
supergravity theories. For example, eleven dimensional
supergravity  compactified on an $n$-torus $T^n$ yields in
$D=11-n$ dimensions a  maximally supersymmetric supergravity
theory with the global symmetry group $E_{n,n}$
\cite{Cremmer:1981zg,Cremmer:1997ct}\footnote{Here $E_{n,n}$ is
the maximal noncompact form of the exceptional group $E_{n}$. For
brevity we write them simply as $E_{n}$. For $n \leq 5$ we have
$E_{0}$ trivial, $E_{1} = \IR, E_{2} = GL(2, \IR), E_{3} = SL(3,
\IR) \times SL(2, \IR), E_{4} = SL(5, \IR), E_{5} = O(5, 5)$.}.
The $SL(n,\IR)$ part of the symmetry group $E_n$ is obvious and is
associated with the large diffeomorphism group of the internal
torus $T^n$. The symmetry enhancement becomes more clear in
string/M theory. The discrete subgroup $E_n(Z)$ is the U-duality
symmetry of M-theory compactified on $T^n$ \cite{Hull:1994ys}. The
$SL(n,Z)$ part is the geometric part of the T-duality group,
whereas the enhancement is associated with the
 part of the T-duality that mixes momentum  and
winding modes and the S-duality (which we will refer to as the
non-geometric symmetries from here on). As we have reviewed here
for $n=2$ case, if the duality twist is in the geometric
$SL(n,\IR)$, then the reduction can be lifted to a string theory
compactification on a twisted torus (can be viewed as having
arisen as the low energy effective theory limit of such a
compactification). On the other hand, when the symmetry is a
non-geometric symmetry, then in most cases there is no lifting to
a conventional geometric string background
\cite{Hull:2004in,Dabholkar:2005ve,Flournoy:2004vn,Shelton:2005cf,Flournoy:2005xe}.
However, the supergravity analysis that we have used here is still
valid \cite{Dabholkar:2002sy,Hull:2003kr,Dabholkar:2005ve} and one
can try and learn from the gauged supergravities arising from
dimensional reductions of supergravity theories with general
duality twists. This is the approach that has been adopted in the
recent paper of Dabholkar and Hull \cite{Dabholkar:2005ve}. If the
duality twist is through an element of the group that contains and
is properly larger than the geometric $SL(n,\IR)$ subgroup then,
from the point of view of the higher dimensional supergravity
theory, this amounts to twisting the internal geometry
\textit{and} introducing non-trivial boundary conditions for some
of the fields as in (\ref{bc}). For this reason, it is important
to study duality-twisted reductions on twisted geometries. The
analysis that we have presented here is a first step in
understanding such dimensional reductions.

As a particular example, we considered here the dimensional
reduction of the eight dimensional type II supergravity theory
(truncated to the sector of $SL(2,\IR)$-singlets) with an
$SL(3,\IR)$ twist on a three dimensional unimodular,
non-semisimple group, which has the local structure of a twisted
torus. This eight dimensional theory can be obtained from a $T^3$
compactification of the eleven dimensional supergravity and the
$SL(3,\IR)$ symmetry that we exploit is the geometric symmetry
associated with the large diffeomorphisms of $T^3$. So, the new
five dimensional gauged supergravity that we found in section 5
can be viewed as having been obtained from a compactification of
M-theory on a non-trivial $T^3$ bundle over a three dimensional
twisted torus (which itself is  a non-trivial $T^2$ bundle over
$S^1$). On the other hand, the same eight dimensional theory that
we start with can also be obtained as the low energy limit of Type
II string theory compactified on a two dimensional torus, $T^2$.
There is an $SL(2,\IR)$ symmetry acting on the K\"{a}hler modulus
of this two-torus (which is the non-geometric part of the
T-duality group). It combines with the S-duality $SL(2,\IR)$ of
Type IIB supergravity to form the $SL(3,\IR)$ symmetry group that
we use to introduce a duality-twisted reduction ansatz. So, from
the point of view of ten dimensional Type IIB supergravity, the
geometry of the five dimensional internal space that leads to the
five dimensional gauged supergravity (\ref{lag5d}) is more
complicated (and perhaps should be analyzed in the context of
Hull's T-folds \cite{Hull:2004in,Hull:2006va}) as the K\"{a}hler
modulus of $T^2$ varies as it traverses over the three dimensional
twisted torus. There are also twisted boundary conditions for
higher dimensional fields determined by their transformations
under the $SL(2,\IR)$ S-duality symmetry.

It would be interesting to study the duality-twisted reductions of
other supergravity theories on more general twisted geometries and
see what this can teach us about string compactifications with
U-duality twists. An important step in this direction would be to
study the moduli space of flat connections on twisted tori.

\section*{Acknowledgements}
I am grateful to Chris Hull for  insightful discussions, for
reading the manuscript and for useful suggestions.  This work is
supported by Irish Research Council for Science, Engineering and
Technology (IRCSET) under the postdoctoral fellowship scheme.

\section*{Appendix}

\appendix

\section{Dimensional Reduction of $\CL_8$}

In this appendix we give the details of the dimensional reduction
of the eight dimensional theory (\ref{8d}) on a twisted torus with
a duality twist in $SL(3,\IR)$.

Let us start by rewriting (\ref{gmfd}), which is the metric ansatz
for our dimensional reduction
\begin{equation}\label{app1}
d\hat{s}^2 = e^{2\alpha \phi} ds^2 + e^{2\beta \phi} \CM_{mn} h^m
h^n
\end{equation}
with $h^m = \eta^m + \CA^m = \eta^m + \CA^m_{\ \mu} dx^{\mu}$,
where $\eta^m$ are the basis one-forms on the three dimensional
group manifold which we presented in
(\ref{eta2}),(\ref{eta3}),(\ref{eta4}) for the cases of our
interest. Above we have split  the eight dimensional curved index
$\hat{\mu}$ as $\hat{\mu}: (\mu, m)$, where $\mu$ runs from 1 to 5
whereas the internal index $m$ runs from 1 to 3. There is a
corresponding splitting of the flat indices $\hat{a}: (a, i)$.
From (\ref{app1}) one can find the vielbeins, that is, the basis
1-forms on the flat tangent space:
\begin{eqnarray}\label{app2}
\hat{e}_{\mu}^{\ a} = e^{\alpha \phi} e_{\mu}^{\ a}, \ \ \ \ \
\hat{e}_{\mu}^{\ a} = 0, \nonumber \\
\hat{e}^i = e^{\beta \phi} e^i = e^{\beta \phi} L_m^{\ i} h^m,
\end{eqnarray}
where we have used that $\CM_{mn}$ is a symmetric matrix so that
it can be written as
\begin{equation}\label{app3}
\CM_{mn} = L_m^{\ i} L_n^{\ j} \delta_{ij}.
\end{equation}
From (\ref{app2}) it follows that
\begin{equation}\label{app4}
\hat{e}_{\mu}^{\ i} = e^{\beta \phi} L_m^{\ i} \CA_{\mu}^{\ m}, \
\ \ \ \ \hat{e}_n^{\ i} = e^{\beta \phi} e_n^{\ i} = e^{\beta
\phi} L_m^{\ i} U^m_{\ \ n}.
\end{equation}
Then we can write
\begin{equation}\label{app5}
\hat{e}_{\hat{\mu}}^{\ \ \hat{a}} = \left(\begin{array}{cc}
                                         e^{-\alpha \phi}
                                         e_{\mu}^{\ a} & e^{\beta
                                         \phi} L_m^{\ i}
                                         \CA_{\mu}^{\ m} \\
                                         0 & e^{\beta \phi} L_n^{\
                                         i} U^n_{\ m}
                                         \end{array}\right).
\end{equation}
Note that $\alpha$ and $\beta$ can be found from (\ref{alfa}) by
using $D=5, d=3$ to be
\begin{equation}\label{app6}
\alpha = \frac{1}{2\sqrt{3}}, \ \ \ \ \beta = - \alpha = -
\frac{1}{2 \sqrt{3}}.
\end{equation}

Performing the dimensional reduction on the flat tangent space is
standard and is discussed in detail in the literature. We refer
the reader to \cite{miriam,Roest:2004pk} for the reduction of the
Einstein-Hilbert term which yields in five dimensions the term
(\ref{EHgmfd}) with $\alpha$ and $\beta$ as in (\ref{app6}). Here
we will discuss the dimensional reduction of the kinetic terms for
the p-form fields and the scalars, for which one also has to
consider the new features due to the non-trivial $SL(3,\IR)$
twist. To analyze this we first need to see how the eight
dimensional Hodge operator is related to the five and three
dimensional ones. One can show, by using (\ref{app2}) and
(\ref{app4}) that
\begin{eqnarray}\label{app7}
\hat{*}(X_{(p)m_1 \cdots m_r} \frac{h^{m_1}\wedge \cdots \wedge
h^{m_{r}}}{r!}) & = & e^{-p \alpha \phi} e^{-r \beta \phi}
e^{(D-p)\alpha \phi} e^{(d-r)\beta \phi} *_D (X_{(p)}) *_d
(\frac{h^{m_1}\wedge \cdots \wedge h^{m_{r}}}{r!}) \nonumber \\
& = & e^{(1-p+r)/\sqrt{3} \phi} *_5 (X_{(p)}) *_3
(\frac{h^{m_1}\wedge \cdots \wedge h^{m_{r}}}{r!}).
\end{eqnarray}
In the last line we have used that $D=5, d=3$ and $\alpha$ and
$\beta$ are as in (\ref{app6}). Then we have
\begin{eqnarray}\label{app8}
\hat{*}_8\hat{H}_{(3)} & = & \hat{*}_8[H_{(3)} + H_{(2)n} \wedge
h^n  + \frac{1}{2!}\epsilon_{mnp} H_{(1)}^m  \wedge h^n \wedge h^p
+ \frac{1}{3!} \epsilon_{mnp} H_{(0)} h^m \wedge h^n \wedge h^p]
\nonumber \\
& = & e^{-4 \alpha \phi} *_5 H_{(3)} *_3 1 + *_5 H_{(2)n} *_3 h^n
+ e^{2 \alpha \phi} \epsilon_{mnp} *_5 H_{(1)}^m *_3 (\frac{h^n
\wedge h^p}{2!}) \nonumber \\
& & + e^{8 \alpha \phi} *_5 H_{(0)} *_3 (\epsilon_{mnp} \frac{h^m
\wedge h^n \wedge h^p}{3!}).
\end{eqnarray}
From (\ref{app8}) we can easily deduce that the dimensional
reduction of $\hat{H}_{(3)}^t \hat{\CM}^{-1} \hat{*}_8
\hat{H}_{(3)}$ yields (\ref{kinetic5d}) by noticing a few
important points we list below.
\begin{equation}\label{app9}
h^m \wedge *_3 h^n = L^m_{\ i} L^n_{\ j} e^i \wedge *_3 e^j =
L^m_{\ i} L^n_{\ j} \delta^{ij} = \CM^{mn} *_3 1.
\end{equation}
$$(\frac{1}{3!} \epsilon_{mnp} h^m \wedge h^n \wedge h^p) \ \wedge
*_3 \ (\frac{1}{3!} \epsilon_{qrs} h^q \wedge h^r \wedge h^s) =
*_3 1.$$
\begin{eqnarray*}
\epsilon_{mpq}\epsilon_{nrs}(\frac{h^p \wedge h^q}{2!}) \wedge *_3
(\frac{h^r \wedge h^s}{2!}) &=& \epsilon_{i_1 i_2
i_3}\epsilon_{j_1 j_2 j_3}(\frac{e^{i_2} \wedge e^{i_3}}{2!})
\wedge
*_3 (\frac{e^{j_2} \wedge e^{j_3}}{2!}) \nonumber \\
& = & \epsilon_{i_1 i_2 i_3}\epsilon_{j_1 j_2 j_3}
\frac{(\delta^{i_2 j_2} \delta^{i_3 j_3} - \delta^{i_3 j_2}
\delta^{i_2 j_3})}{4} *_3 1 = \frac{\epsilon_{i_1 j_2 j_3}
\epsilon_{j_1 j_2 j_3}}{2} *_3 1 \nonumber \\
& = & \frac{(\delta_{j_2 j_2} \delta_{i_1 j_1} - \delta_{j_2 i_1}
\delta_{j_1 j_2})}{2} *_3 1 = \delta_{i_1 j_1}.
\end{eqnarray*}

Now consider the dimensional reduction for the kinetic term
(\ref{kinetics}) for the scalar fields. Following (\ref{yenibir})
we find
\begin{eqnarray}\label{app10}
d\hat{\CK} & = & \tilde{g}(M_p \CK
\eta^p + d\CK + \CK M_p^t \eta^p) \tilde{g}^t \nonumber \\
& = & \tilde{g}(M_p \CK h^p - M_p \CK \CA^p + d\CK + \CK M_p^t h^p
- \CK M_p^t \CA^p)\tilde{g}^t \nonumber \\
& = & \tilde{g}(M_i \CK e^i - M_p \CK \CA^p + d\CK + \CK M_i^t e^i
- \CK M_p^t \CA^p)\tilde{g}^t,
\end{eqnarray}
where $M_i = L_i^{\ m} M_m$. (Remember that $e^i = L^i_{\ m}
h^m$.) Similarly,
\begin{eqnarray}\label{app11}
d\hat{\CK}^{-1} &=& (\tilde{g}^t)^{-1}(-M_p^t \CK^{-1} \eta^p +
d\CK^{-1} - \CK^{-1} M_p \eta^p)\tilde{g}^{-1} \nonumber \\
& = & (\tilde{g}^t)^{-1}(-M_p^t \CK^{-1} h^p + M_p^t \CK^{-1}
\CA^p + d\CK^{-1} - \CK^{-1} M_p h^p + \CK^{-1} M_p
\CA^p)\tilde{g}^{-1} \nonumber \\
&=&(\tilde{g}^t)^{-1}(-M_i^t \CK^{-1} e^i + M_p^t \CK^{-1} \CA^p +
d\CK^{-1} - \CK^{-1} M_i e^i + \CK^{-1} M_p \CA^p)\tilde{g}^{-1}
\end{eqnarray}
From (\ref{app10}) and (\ref{app11}) we have
\begin{equation}\label{app12}
{\rm tr}(d\hat{\CK}\wedge \hat{*}d\hat{\CK}^{-1}) = {\rm
tr}[\tilde{g}(D\CK + (M_i \CK + \CK M_i^t)e^i) \tilde{g}^t \wedge
*_3 (\tilde{g}^t)^{-1}(D\CK^{-1} -(M_i^t \CK^{-1} + \CK^{-1} M_i)
e^i) \tilde{g}^{-1}],
\end{equation}
where $D\CK$ and $D\CK^{-1}$ are as in (\ref{yeniuc}). From
(\ref{app12}) we find (\ref{scalar5d}) by using (\ref{app7}) and
$e^i \wedge *_3 e^j = \delta^{ij} *_3 1$.

Let us now look at the dimensional reduction of the topological
term
\begin{equation}\label{app13}
\epsilon^{abc} \hat{H}_{(3)a} \wedge \hat{H}_{(3)b} \wedge
\hat{B}_{(2)c},
\end{equation}
where $\hat{B}_{(2)}$ and $\hat{H}_{(3)}$ are as in (\ref{Bfield})
and (\ref{H3}), respectively. Note that the $y$ dependence
introduced through $\tilde{g}(y)$ cancels out in five dimensions
as we have
$$\epsilon^{abc} S_a^{\ d} S_b^{\ e} S_c^{\ f} = \epsilon^{def},$$
for all $S \in SL(3)$. The term (\ref{app13}) does not involve the
metric, so its dimensional reduction is straightforward. The
important point we should note is that the only non-zero terms are
the ones of the form
$$ \sim \epsilon H_{(i)} \wedge H_{(j)} \wedge B_{(k)} \wedge h
\wedge h \wedge h, $$ with $i+j+k=5$ as the dimensions of the
internal and the base space are three and five respectively. From
the product (\ref{app13}) one obtains nine such terms some of
which cancel each other. It is easy to check that the remaining
terms give (\ref{cs5d}).


\begin{thebibliography}{99}

\bibitem{SS1} J.~Scherk and J.~H.~Schwarz, ``Spontaneous Breaking Of
Supersymmetry Through Dimensional Reduction,'' Phys.\ Lett.\ B
{\bf 82} (1979) 60.

\bibitem{Dabholkar:2002sy}
  A.~Dabholkar and C.~Hull,
  ``Duality twists, orbifolds, and fluxes,''
  JHEP {\bf 0309} (2003) 054
  [arXiv:hep-th/0210209].


\bibitem{SS2} J.~Scherk and J.~H.~Schwarz, ``How to get masses from
extra dimensions", Nucl.\ Phys.\ B {\bf 153}, (1979) 61 .

\bibitem{Hull:2005hk}
  C.~M.~Hull and R.~A.~Reid-Edwards,
  ``Flux compactifications of string theory on twisted tori,''
  arXiv:hep-th/0503114.


\bibitem{miriam}
M.~Cvetic, G.~W.~Gibbons, H.~Lu and C.~N.~Pope, ``Consistent group
and coset reductions of the bosonic string,''
arXiv:hep-th/0306043.

\bibitem{Roest:2004pk}
  D.~Roest,
  ``M-theory and gauged supergravities,''
  Fortsch.\ Phys.\  {\bf 53} (2005) 119
  [arXiv:hep-th/0408175].


\bibitem{Dall'Agata:2005ff}
  G.~Dall'Agata and S.~Ferrara,
  ``Gauged supergravity algebras from twisted tori compactifications with
  fluxes,''
  Nucl.\ Phys.\ B {\bf 717} (2005) 223
  [arXiv:hep-th/0502066].


\bibitem{myers1}
N.~Kaloper and R.~C.~Myers, ``The O(dd) story of massive
supergravity,'' JHEP {\bf 9905}, (1999) 010
[arXiv:hep-th/9901045].

\bibitem{Bergshoeff:2003ri}
  E.~Bergshoeff, U.~Gran, R.~Linares, M.~Nielsen, T.~Ortin and D.~Roest,
  ``The Bianchi classification of maximal D = 8 gauged supergravities,''
  Class.\ Quant.\ Grav.\  {\bf 20} (2003) 3997
  [arXiv:hep-th/0306179].



\bibitem{Hull:1994ys}
  C.~M.~Hull and P.~K.~Townsend,
  ``Unity of superstring dualities,''
  Nucl.\ Phys.\ B {\bf 438} (1995) 109
  [arXiv:hep-th/9410167].


\bibitem{salam}
A.~Salam and E.~Sezgin, ``D = 8 Supergravity,'' Nucl.\ Phys.\ B
{\bf 258} (1985) 284.


\bibitem{Alonso-Alberca:2000gh}
  N.~Alonso-Alberca, P.~Meessen and T.~Ortin,
  ``An SL(3,Z) multiplet of 8-dimensional type II supergravity theories and
  the gauged supergravity inside,''
  Nucl.\ Phys.\ B {\bf 602} (2001) 329
  [arXiv:hep-th/0012032].


\bibitem{Cremmer:1981zg}
E.~Cremmer, ``Dimensional Reduction In Field Theory And Hidden
Symmetries In Extended Supergravity,'' LPTENS 81/18
{\it Lectures given at ICTP Spring School Supergravity, Trieste,
Italy, Apr 22 - May 6, 1981}



\bibitem{Cremmer:1997ct}
  E.~Cremmer, B.~Julia, H.~Lu and C.~N.~Pope,
  ``Dualisation of dualities. I,''
  Nucl.\ Phys.\ B {\bf 523} (1998) 73
  [arXiv:hep-th/9710119].




\bibitem{Hull:2004in}
  C.~M.~Hull,
  ``A geometry for non-geometric string backgrounds,''
  JHEP {\bf 0510} (2005) 065
  [arXiv:hep-th/0406102].






\bibitem{Dabholkar:2005ve}
  A.~Dabholkar and C.~Hull,
  ``Generalised T-duality and non-geometric backgrounds,''
  JHEP {\bf 0605} (2006) 009
  [arXiv:hep-th/0512005].



\bibitem{Flournoy:2004vn}
  A.~Flournoy, B.~Wecht and B.~Williams,
  ``Constructing nongeometric vacua in string theory,''
  Nucl.\ Phys.\ B {\bf 706} (2005) 127
  [arXiv:hep-th/0404217].


\bibitem{Shelton:2005cf}
  J.~Shelton, W.~Taylor and B.~Wecht,
  ``Nongeometric flux compactifications,''
  JHEP {\bf 0510} (2005) 085
  [arXiv:hep-th/0508133].





\bibitem{Flournoy:2005xe}
  A.~Flournoy and B.~Williams,
  ``Nongeometry, duality twists, and the worldsheet,''
  JHEP {\bf 0601} (2006) 166
  [arXiv:hep-th/0511126].



\bibitem{Hull:2003kr}
  C.~M.~Hull and A.~Catal-Ozer,
  ``Compactifications with S-duality twists,''
  JHEP {\bf 0310} (2003) 034
  [arXiv:hep-th/0308133].


\bibitem{Hull:2006va}
  C.~M.~Hull,
  ``Doubled geometry and T-folds,''
  arXiv:hep-th/0605149.




\end{thebibliography}
\end{document}